# The Ecosystem of Trust (EoT): Enabling effective deployment of autonomous systems through collaborative and trusted ecosystems

**Authors**: Jon Arne Glomsrud[1], Tita Alissa Bach[1,2] (ORCID: 0000-0001-8536-6826)

[1]Group Research and Development, DNV, Høvik, Norway

[2]Tita.Alissa.Bach@dnv.com

**Acknowledgment**
We would like to thank Dr. Øyvind Smogeli for his critical review of the earlier version of this manuscript.

**Statements and Declarations**
Authors have declared no conflict of interest.

**Funding**
This work is partly sponsored by the Research Council of Norway through the project TRUSST with project number 313921.

**Author contributions**

JAG contributed to the conception of the work. TAB matured the idea. Both authors drafted and critically revised the article in an iterative process. Both authors approved the final version.

**Abstract**

Ecosystems are ubiquitous, but trust within them is not guaranteed. Trust is paramount because stakeholders within an ecosystem must collaborate to achieve their objectives. With the twin transitions (digital transformation to go in parallel with green transition) accelerating the deployment of autonomous systems, trust has become even more critical to ensure that the deployed technology creates value. To address this need, we propose an ecosystem of trust (EoT) approach to support deployment of technology by enabling trust among and between stakeholders, technologies and infrastructures, institutions and governance, and the artificial and natural environments in an ecosystem. The approach can help the ecosystem's stakeholders to create, deliver, and receive value by addressing their concerns and aligning their objectives. We present an autonomous, zero-emission ferry as a real-world use case to demonstrate the approach from a stakeholder perspective. We argue that assurance (grounds for justified confidence originated from evidence and knowledge) is a prerequisite to enable the approach. Assurance provides evidence and knowledge that are collected, analysed, and communicated in a systematic, targeted, and meaningful way. Assurance can enable the approach to help successfully deploy technology by ensuring that risk is managed, trust is shared, and value is created.

# 1. Introduction

## 1.1. Ecosystems

"Ecosystems" exist everywhere we go. The term was first defined by botanist A. G. Tansley (1935) to describe the idea of "the whole physical system including the whole complex of physical factors forming an environment of the biome, the habitat factors in the widest sense, in which organisms cannot be separated from their special environment, with which they form one physical system" (p. 299). Since then, the term has been used widely beyond the fields of biology in an arbitrary and non-specific manner. For example, it has been combined with sustainable goal descriptions (Auerswald & Dani, 2018; Joo & Shin, 2018; Ma et al., 2018; Romanelli, 2018; Shi et al., 2021) and used to describe economic, business, digital (van de Hoven et al., 2021), socio-technical, and socio-ecological (Ahlborg et al., 2019) situations. In brief, an ecosystem is a complex and interdependent system.

We propose an *Ecosystem of Trust* (EoT) as a human-centric approach to understand and shape an ecosystem by fostering trust among and between stakeholders, technologies and infrastructures, institutions and governance, and the artificial and natural environments *that depend on each other*, so that the ecosystem's stakeholders and other living creatures can create, deliver, and receive value.

In this description, stakeholders are people, communities, and organisations; technologies and infrastructures can be digital technologies, automation, and physical systems; institutions and governance represent any structure of law, regulation, culture, and/or norms; and all living creatures include humans, plants, and animals. These interdependencies and interactions between entities in an ecosystem are fundamental to the EoT approach; if a change impacts one entity, it has an impact on others as well. Consequently, trust between these entities is a prerequisite for an ecosystem to function. It is a concept so central and profound, that without trust, chaos is guaranteed.

Ecosystems typically form organically, and trust may not necessarily be present between all entities. For example, a pack of wolves in a wildlife ecosystem often trust each other but they may not trust wolves from other packs (Zimen, 1976). Similarly, in an ecosystem with human beings, trust may not be present between all stakeholders, elements, and systems, preventing some value being created. The absence of trust can become a problem because all entities within an ecosystem need to work together so that its stakeholders and other living creatures can achieve their objectives. The EoT approach is even more important when a new technology is being deployed because deployment changes many of the dynamics of the ecosystem in question, especially the trust needs of the stakeholders (Hopster, 2021).

## 1.2. Deployment of technology into ecosystems

We focus on the deployment of technology in this paper because technology has been used to create value for society throughout history. This trend has accelerated over the past century. The computer revolution began in the late 1940's (Randell, 1976) and since then autonomous systems (i.e., technologies with autonomy capabilities such as using Artificial Intelligence (AI) that operate with zero or limited human intervention) have been increasingly used to help achieve sustainability and tackle environmental challenges (i.e., twin transitions). These twin transitions (Stefan et al., 2022), i.e., the idea that the digital transformation needs to go in parallel with green transition, of making our society more sustainable (Mondejar et al., 2021) has led to global sustainability efforts including reducing carbon footprints, using limited resources more responsibly, and catering to equality and an equal quality of life. Accordingly, autonomous systems have been used to make industries more sustainable in, for example, food supply chains (Duong et al., 2020), farming, forestry, and the marine/aquaculture sectors (Galaz et al., 2021), fish-farming (*Aquabyte*), renewable energy (Ahmad et al., 2021), home care (Yang et al., 2020), and healthcare (Greco et al., 2020).

Although the development of autonomous systems has been advancing for decades now (Janssen et al., 2019), deploying them for the benefits of society has never been a straight-forward process (Bach et al., 2022; DNV Group Research and Development, 2022; Yigitcanlar et al., 2022). This is likely because autonomous systems, which are by nature complex, are being deployed in unstructured or unpredictable environments (Harel et al., 2020). Reckless deployment has led to disastrous negative consequences such as discrimination and death (Buolamwini & Gebru, 2018; Buolamwini, 2017; Dastin, 2018; Hoffman & Podgurski, 2019; Kayser-Bril, 2020; Kohli & Chadha, 2020; Olteanu et al., 2019; Ruiz, 2019).

Even if autonomous system technologies can be perfected, deploying them into society is a separate and risky endeavour with many different potential outcomes (Sherry et al., 2020). Deployment challenges thus have been the main topic of research in, for example, the human-computer interaction (HCI) field (e.g. Bellet et al., 2019; Carmona et al., 2021). This is because autonomous system development usually focuses on building robust technology, whereas deployment is heavily dependent on human factors and the ecosystem where the interactions between humans and technology happen.

These deployment risks have raised trust issues for their stakeholders (e.g., end-users, regulators, and operators) (Bach et al., 2022): Are the technologies safe? Can we trust the manufacturers? Is the technology compliant to relevant regulations? Are the regulations fit-for-purpose? Who is held accountable should there be an accident? Are there unforeseen consequences of using the technologies? Will deploying a technology disrupt the ecology? Twin transitions have accelerated the deployment of autonomous systems, amplifying these trust issues.

Deploying new technologies challenges humans' ability to understand them and decide whether to trust and use them (Bach et al., 2022). Unlocking the potential of technologies ideally needs all relevant stakeholders to trust the technologies, the process, and each other, or the technologies will not be used optimally. In addition, there are concerns that deploying any technology, let alone an autonomous system, is likely to change the dynamics of the ecosystem where it is deployed (Hopster, 2021). These concerns are linked to whether the technologies will create value as intended with minimal negative consequences.

Addressing these concerns and fostering stakeholder trust requires a collective effort to reduce the likelihood and severity of potentially negative outcomes related to use of technology, while ensuring that the intended value and purpose are created and achieved by the technology being deploying in the respective ecosystem. The EoT approach is valuable in understanding how trust can be fostered by aligning and meeting stakeholders' different trust needs, so that participation in the EoT can be made more attractive. This is because not every stakeholder's objective in an ecosystem is directed towards achieving sustainability. Nevertheless, their participation in fostering trust in the ecosystem is key to success for twin transitions as sustainability can only be achieved if everybody participates and contributes (Mondejar et al., 2021; Stefan et al., 2022). Importantly, the EoT approach can help the ecosystem's stakeholders to pursue their own goals without the expense of others.

### 1.3. Deploying technology using the EoT approach

In this paper, the EoT approach that we are proposing focuses on understanding and onboarding stakeholders to participate in the trustworthy and responsible practices and interactions so that trust is present among and between all stakeholders and technology. Following the twin transitions' perspective, every deployed technology should have the goal to contribute to achieving sustainability (**Fig. 1**). Participating in trustworthy and responsible practices and interactions is essential to shaping an ecosystem where stakeholders with different interests and objectives come together to align and pursue common good (e.g., sustainability). Therefore, attracting stakeholders to participate is a fundamental part of the approach.

Onboarding stakeholders to participate in trustworthy and responsible practices and interactions should happen from the early phase of development of technology, or even earlier, to understand the dynamics of the ecosystem where the technology is to be deployed (**Fig. 1**). This is useful because deploying the same technology can have different outcomes in different ecosystems. Understanding the ecosystem while in the development phase can help to develop technology that is fit-for-purpose for the ecosystem in question.

For this reason, the approach can encourage stakeholders to identify and address each other's trust needs, while aligning each other's objectives to meet the common good. Specifically, applying the approach can identify patterns of interdependency so that stakeholders can recognise ripple effects of a change to one group of stakeholders on others (e.g., stakeholders, elements, or systems). This is important because a change can affect the others in different ways. Tracking these changes eventually creates a complete picture of what and how the event of change affects the ecosystem.

Additionally, these interdependencies are links that can identify some stakeholders or entities that may not necessarily be identified otherwise (e.g., more vulnerable or minority stakeholders, other overlooked technologies). Another potential result of applying the approach is unpacking different stakeholders' motivation, interests, and concerns that drive or prevent interactions among or between stakeholders and technologies. Understanding the dynamics of an ecosystem in greater detail can help to create a deployment strategy that is more likely to succeed. The strategy should cover pre- and post- deployment stages to ensure that what is planned is what happens.

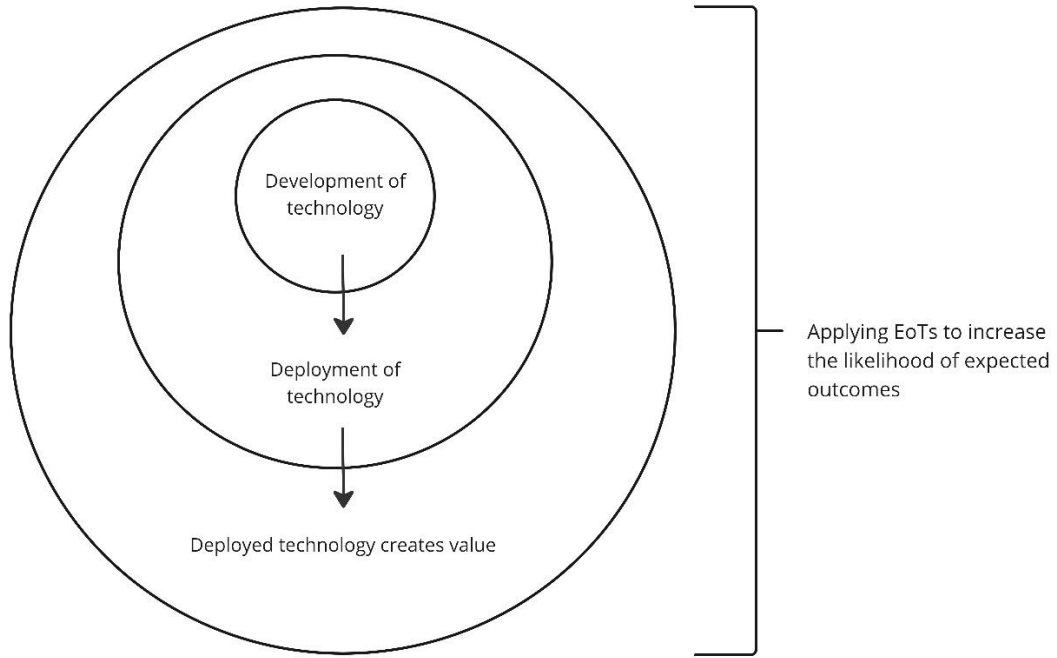

**Fig. 1** The EoT approach is ideally applied from the development stage through to the deployment of technology to increase the likelihood that technology being developed creates value as expected with minimal negative consequences as it is being deployed

### 1.4. Contribution of the EoT approach

Deploying autonomous systems naturally draws interdisciplinary approaches. Thus, the EoT approach can enable cross-collaborations between industry and academia and private and public sectors from ideation, to development, through to deployment and continuous monitoring of autonomous systems. The approach can also be used to guide research, industry, policymakers, and individuals to collectively deploy any technology in question successfully, unlocking its value for society at large.

We are using a specific use case of an autonomous, zero-emission ferry for urban mobility to demonstrate how our approach can be used in practice. As an industry actor focused on providing assurance, we were involved in maturing the assurance approach to be fit-for-purpose for the specific ferry (Pantelatos et al., 2023). The maturing process was completed in the context of Sustainable Development Goals (SDGs) as well as focused on engaging the local population as potential passengers (Pantelatos & St.Clair, 2022; *The TRUSST project*). The maturing process covered the project from concept, to the creation of the physical ferry, to understanding the context where the ferry would operate (Jacobsen & St.Clair, 2023).

Our approach can not only be applied to helping deploy such ferries successfully, but also to other autonomous systems specifically, and any technology in general. In a broader sense, we find that understanding how an ecosystem works and creating trust in an ecosystem is important for laying a foundation that can be developed further into the ecosystem that works for all.

Finally, the approach can be used for exploring and discussing privacy issues within the ecosystem where digital technologies are being deployed and interconnected. Interconnected digital technologies are relatively more vulnerable against cyberattacks that may reveal personal data. In addition, data coming from various sources within the same ecosystem may become identifiable. The EoT approach can identify these interdependencies and potential privacy issues for all stakeholders. In our use case, while it may not be obvious why the deployment of a passenger ferry would draw attention to privacy-related issues, it is important to note that the ferry will be deployed for public use in a public area. Autonomous systems require and generate massive amount of data to operate as intended through various channels. The fact that even anonymous electronic bus tickets can reveal passengers' identity in a community (Avoine et al., 2014) raises questions regarding what personal data might be collected even unintentionally by operating the autonomous ferry that employs interconnected communication systems (Amro & Gkioulos, 2022).

## 2. Use case: The autonomous, zero-emission ferry

The concept of an autonomous, zero-emission ferry was born through several concurrent events. A professor at a leading university in Trondheim (i.e., Norwegian University of Science and Technology), the $4^{th}$ biggest city in Norway, opposed a new footbridge that would block the vintage boats that travelled the canal. At the same time, the city's inhabitants had similar opposition to road and bridge building when the Trondheim's municipality planned to build one more bridge in Ravnkloa. These desires aligned and the idea of a ferry that could solve the city's density and congestion issues while increasing the mobility of the inhabitants sounded very appealing.

This led to the development of an autonomous, zero-emission ferry instead of another bridge, road, or tunnel. It also inspired the creation of a new research program called Autoferry at the university, (*Autonomous all-electric passenger ferries for urban water transport (Autoferry)*), a new start-up to realise the idea (*Zeabuz*), a partnership between technology providers, regulators, and an assurance provider (*Autonomous urban mobility: Taking digital assurance to the next level*; *The TRUSST project*), and engagement with the local population (*Assuring Trustworthy, Safe and Sustainable Transport for All*).

Jacobsen and St. Clair explain in detail the path the ferry took to become the radical twin innovations in their conference presentation (Jacobsen & St.Clair, 2023). Deploying the ferry changed the perception of the water as a blockage for transport and mobility to one that could open new living spaces that were previously inaccessible. Another ferry based on this design is planned for operation in Stockholm and is currently in its testing stage with the plan of removing a human operator onboard over time (*Torghatten to launch autonomous ferry in Stockholm* ). Although neither ferry is yet fully autonomous, we extrapolate the ferry concept into a full autonomous one to follow the ultimate intended purpose of the creation of the ferry.

### 2.1. The stakeholders of the ferry

The benefits of the ferry cannot be realised if the passengers do not trust it and refuse to use it. This scenario can lead to abandonment of the project and the loss of the potential societal value. We use this autonomous, zero-emission ferry as a use case because the potential ripple effect in the ecosystem where it is being deployed demonstrates the importance of the EoT approach.

In a case of deploying a novel autonomous vehicle into a local community or a city, the EoT approach can be used as a point of departure to ensure a holistic approach covering all stakeholders, trust needs and concerns, technologies, elements, and systems. For example, it can ensure that no stakeholders gain benefits at the expense of others (e.g., business-minded stakeholders operating an unsafe ferry at the expense of the passengers to gain profit by cutting manufacturing cost), and to align all stakeholders' objectives towards achieving sustainability.

We use the passengers' point of view throughout the use case in this paper because passengers are the key stakeholder group that holds the highest risk and are the end users who can stop or enable such technology. The ecosystem in our use case of the autonomous, zero-emission ferry consists of several groups of stakeholders categorised by the level of impact a stakeholder will experience if an incident or a near miss happens to the ferry. Below is a simplified stakeholder categorisation:

(i) First-hand stakeholders: The first group is those interacting directly with the ferry. This includes passengers, onshore or remote operators, other actors in the water such as other waterborne vehicles, swimmers, and kayakers. This paper emphasises this stakeholder group's point of view because passengers carry the highest risk in terms of the probability and severity of damage in the event of an incident or a near miss. Over time, when the ferry creates new travelling patterns and changes the overall transportation system of the city, this group will expand to include the inhabitants of the city and its surroundings.

(ii) Second-hand stakeholders: The second group is those who are involved in the manufacturing and development of the technology including software and product developers, technicians and engineers, and vendors or suppliers.

(iii) Third-hand stakeholders: The third group is those who invest in the ferry and its infrastructure and expect success of the ferry. This group includes investors, regulators, insurance companies, and other funders. The creation of new travelling patterns will expand this group to include city governments, real-estate business and development, and the construction industry.

In an ideal scenario, the ferry will create a new ecosystem that provides its first-hand stakeholders with efficient and sustainable transport that businesses can profit from and a more sustainable city with less (ideally no) environmental impact.

We use an illustration shown in **Fig. 2** to describe two opposing scenarios of how the ferry could change the patterns of traveling and behaviours of the inhabitants.

*The first scenario* describes when the EoT approach is applied pre- and post- deployment of the ferry. In this case, the inhabitants and buildings on the left side are isolated prior to the deployment of the ferry because the land is at the end of a peninsula. The inhabitants must drive around the water to reach the right side where the city centre and most business offices are. The inhabitants on the left start to have local discussions that a ferry crossing this waterway would be a useful solution and have little environmental impact.

This idea from the inhabitants is quickly picked up by the leading university in the city, who immediately involve regulators and other relevant stakeholders. Stakeholder identification and involvement are optimised using the EoT approach to identify who and what will be affected by the deployment of such a ferry. Based on the findings, the ferry is developed in an interdisciplinary environment. As a result, they develop a ferry that is autonomous and produces zero emissions because, based on the EoT analysis, such a ferry will have little to no environmental impact but will meet the needs of the population.

Once the ferry is developed and approved by regulators, a deployment plan is created and finalised by all stakeholders and the manufacturer, giving everyone involved a sense of ownership. As a result, the deployment goes as planned, and the acceptance of the ferry into the existing ecosystem is high.

Deploying the ferry cuts the commuting time for the inhabitants on the left by half. More housing development happens on the left side. New offices also spring up on the left side because inhabitants who live in the city centre can now also easily cross to the left. The city centre expands automatically. In addition, the congestion on the road towards the city centre diminishes and there are no more demands for bridge building. All these changes also lead to reduced pollution for the city.

As the demand for the ferry increases, the ferry doubles its capacity and plans to have more ferries with more frequent departures which benefits businesses. There are also plans to deploy the ferry in other cities and countries surrounded by waterways, creating new business and employment that were not available before. New value is created by the deployment, leveraging existing value from pre-deployment.

This first scenario and all the associated benefits cannot happen if the inhabitants on the left do not trust the ferry.

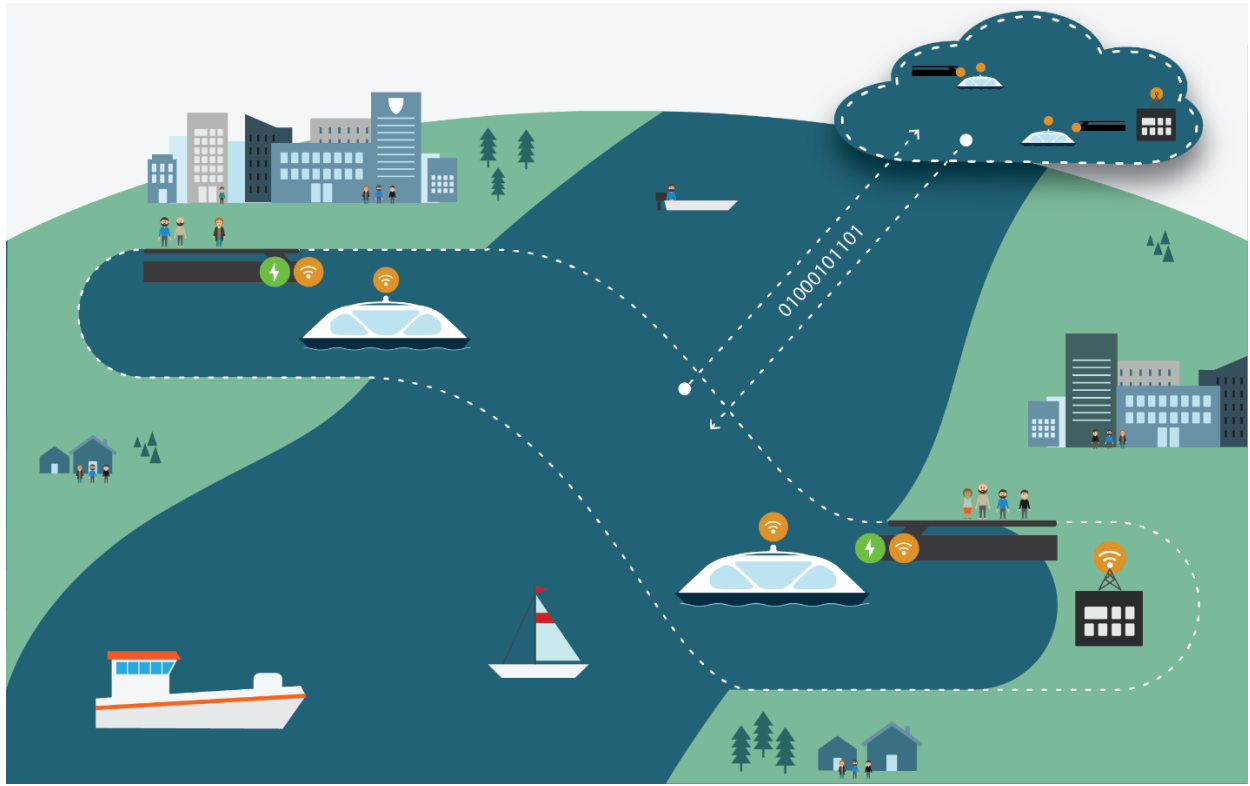

**Fig. 2** Illustrated ecosystem of the ferry that includes passengers waiting in two harbours separated by a waterway, other waterborne vehicles, buildings that are connected by the ferry, and the interconnected technology to operate the ferry (*The TRUSST project*)

In *the second scenario* there is no trust between stakeholders in the ecosystem. The development of the ferry happens in isolation by several technical people at a university. The regulators and potential passengers only become involved late in the process. This results in scepticism when the ferry is almost finished developed. The deployment plan must be postponed significantly to onboard the regulators and inhabitants to the manufacturers' claim that the ferry is in fact safe for operation. Although the inhabitants on the left understand the benefits to them, they are so used to driving around the peninsula that they do not see the ferry as essential.

When the ferry is deployed without major stakeholder involvement, only a few inhabitants are willing to be passengers. The rest do not want to change their behaviour. Especially since the home office has become more common, they cannot see the benefit of using the ferry. With few users, the ferry is eventually abandoned, and its operation is stopped altogether due to the ferry company losing investors and passengers. The dynamic between the inhabitants, regulators, and the university worsens due to the failed ferry project which creates tension and makes future projects even more difficult. Distrust (i.e., a belief that someone or something is untrustworthy) and mistrust (i.e., scepticism about the trustworthiness of something or someone) that were not there prior to the ferry project, have developed between the stakeholder groups.

## 3. The EoT approach fostering "the correct trust"

In the two opposing scenarios above, stakeholder involvement is key to gaining acceptance of new technology and to prevent mistrust. Deploying a technology can have positive or negative consequences. This is because an ecosystem is "alive" and its dynamic can easily change by a change in the interaction between a few stakeholders, such as in the second scenario where the tension is created after failure in deploying the ferry.

Interactions that happen in an ecosystem usually have a sole purpose to create value for the stakeholders specifically and the ecosystem in a broader sense. For example, an interaction might reduce commuting time, develop new businesses, or cross the waterway without impacting the nature and wildlife. These interactions happen because there is an expectation that the interaction will create the expected value. Expectations or dependence on the value from using the technology implies the presence of risk and the need for trust. The receiver of the value thus becomes the trustor, and the one who delivers the value becomes the trustee.

Mayers et al. (2006) define trust as the willingness of a party (the trustor) to be vulnerable to the actions of another party based on the expectation that the other (the trustee) will perform a particular action important to the trustor (i.e., to offer a value), irrespective of the ability to monitor or control that other party. Trust is important because, according to Taddeo (2017), trust is a facilitator of interactions among stakeholders and with technology in an ecosystem.

Trust is thus not the interaction itself but a property of interactions that initiates, enables, or changes the way the interactions occur. For example, if one party does not trust another party, they will be reluctant to initiate interactions. In contrast, if these two parties trust each other, interactions are automatically enabled between them. Undertrusting or overtrusting are equally problematic for an ecosystem.

The absence of trust or undertrusting, which can come in the form of mistrusting or distrusting (Citrin & Stoker, 2018), is likely to lead to misuse and/or disuse. For example, a passenger undertrusting an autonomous ferry can lead to the second scenario as described above. Whereas undertrusting between stakeholders in an ecosystem is likely to lead to strict supervision and control between them, creating tension and inefficiencies.

Although undertrusting can lead to failure of the ferry deployment, we argue that it is a bigger issue in an ecosystem for stakeholders to "trust incorrectly" (Aroyo et al., 2021; Jacovi et al., 2021; Taddeo, 2017), a phrase borrowed from Taddeo (2017). This is because undertrusting is relatively predictable and expected first human reaction when introduced to something new or unfamiliar, and thus we know to make an effort to prevent it. In contrast, the consequences of overtrusting or blindly trusting are not necessarily as well discussed (Aroyo et al., 2021).

Digital technologies, including autonomous systems, have become so embedded in everyday life that their presence is not necessarily visible and often overlooked. At first glance, it may seem implausible to overtrust the autonomous, zero-emission ferry since there is no operator onboard. However, once the ferry is deployed and embedded in an ecosystem and inhabitants' routines, automation complacency and bias can easily lead to overtrusting (Aroyo et al., 2021; Dixon, 2020; Rodriguez et al., 2019). Overtrusting the ferry can look like an operator who is over reliant of the technology of the ferry and skips a boarding safety checklist. Overtrusting between stakeholders is likely to lead to them assuming that everything is fine instead of checking the reality and following safety standards.

Trusting correctly (Taddeo, 2017) within an ecosystem should be based on *assurance*: grounds for justified confidence originated from evidence, knowledge, experiences and/or skills (St.Clair, 2022). The challenge for an ecosystem is how to ensure that the correct trust is present at all times, especially during interactions among and between stakeholders and technology in the ecosystem (Taddeo, 2017). In brief, trusting something or someone correctly means that although we trust someone or something in an ecosystem, we still diligently follow assurance mechanisms such as rules, norms, good practice, standards, and relevant guidance (Scott, 2008). This is the correct trust that the EoT approach focuses on fostering.

Below is an example to illustrate this (**Fig. 3**). Note that we use fictional characters as placeholders in discussing our approach and they do not refer to interpersonal trust. They are used for a convenience purpose and to aid comprehension of the EoT approach.

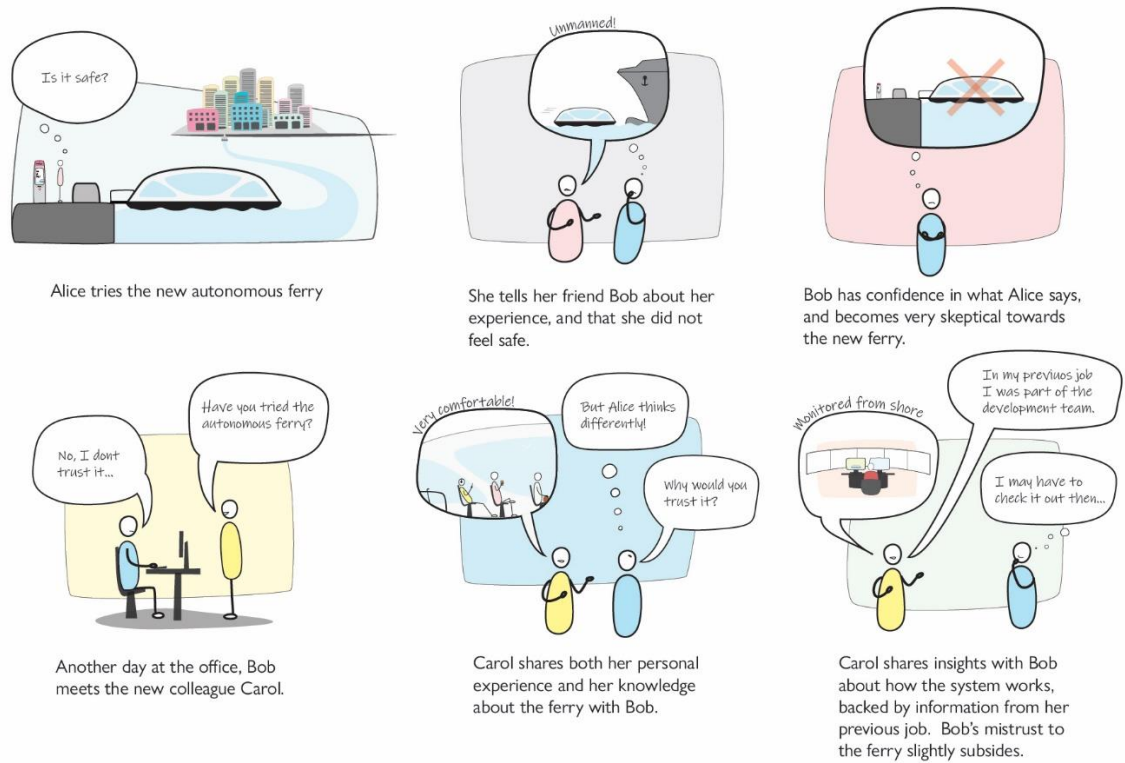

**Fig. 3** Alice, Bob, and Carol on trusting the autonomous, zero-emission ferry correctly

Let us refer to **Fig. 2** again, this time using the first scenario where the deployment goes smoothly, but from the perspective of late adopters (see **Fig. 3** for the summary). Alice, a long-term friend of Bob, tells Bob about her experience being a passenger on the new autonomous ferry that has just been deployed that week in their city. Both Alice and Bob live at the end of the peninsula, and they have a long drive around the water to get to the city centre, to work, shop, etc. Bob and Alice have heard about the ferry due to many stakeholder engagement projects in the city. However, neither of them has been directly involved as they are not technology-driven people and are sceptical of novel technologies such as autonomous systems.

After many conversations, Alice's family and close friends convince her to try out the ferry. Bob has been planning to use the ferry as well because using the ferry will cut the length of his commute to work by more than half. However, Bob is worried about the safety risks and is reluctant to try the ferry.

Alice tells Bob that based on her experience, the new ferry does not make her feel safe to be a passenger. She explains that there is nobody onboard controlling the ferry, she is not familiar with the safety mechanisms being presented through the loudspeakers and the posters, and she is rather anxious when the waves or another ferry come. Hearing Alice's experience, Bob concludes relatively quickly that he has now no intention to try the ferry, even without ever being a passenger on it, and perceives that the risk is not worth the shorter commute.

Bob agreeing with Alice is not blind trust but based on his confidence in Alice's judgement due to his long-term friendship. Bob is familiar with Alice's attitudes, values, behaviours, and her motivation with respect to Bob. Bob has enough grounds to justify his confidence to know that Alice's judgement is trustworthy and sharing her experience is in Bob's interest.

The same week, Bob meets a new colleague, Carol, who also has been a passenger of the new autonomous ferry. Carol shares that the ferry is very safe and even enjoyable for passengers. Bob remembers Alice's experience and does not immediately trust that the ferry is safe, as claimed by Carol. This is because Bob has no experience or history with Carol so that he first needs to establish some level of confidence in any information coming from Carol.

Now, Bob challenges Carol about her experience feeling safe on the ferry as Carol's experience is contradictory to Alice's experience. Bob soon learns that in Carol's previous job, she was involved in the development of the autonomous ferry. She presents evidence and reasoning of the ferry's safety mechanisms in detail with Bob. This discussion leads to Bob's mistrust to the ferry subsiding.

By the end of the discussion, Bob is convinced to check out the ferry. He has made a significant change from not willing to try it at all to considering giving the ferry a chance. Bob's change of heart is not a blind trust, but rather a result of the fact that he now has enough evidence to not immediately mistrust the ferry. Carol being an expert who is also a passenger, the evidence presented, and direct explanations of the safety mechanisms of the ferry have increased Bob's knowledge about Carol's trustworthiness and credibility as well as Bob's knowledge about the ferry. Bob's need of assurance about the ferry is fulfilled.

## 4. Trusting correctly is based on confidence

We have stated that trusting correctly means following relevant regulations, rules, norms, standards, and approaches meant to provide grounds for confidence even though we may have already trusted something or someone. We emphasise that confidence is different from trust. Trust typically (1) involves judgement coming from various factors including cognition, emotions, and motivations integrated into a broader picture (e.g., I trust that the ferry can bring me to the other side safe and sound), thus (2) relies on a broad range of information (e.g., what the other passengers have said, the ferry's communicated safety mechanisms, and own feelings and risk perceptions), and (3) is only meaningful when risks are present (e.g., I know that I can drown if the ferry has an accident) (Adams, 2005).

In addition, Adams makes a distinction that confidence is different from trust because confidence is (1) built on past observations with limited extrapolation, reflections or learning from them (e.g., I am confident that the ferry will be on time today because the ferry was on time yesterday), (2) is specific to a task (e.g., whether the ferry will be on time today), (3) relies only on someone's cognition (e.g., I think the ferry will be on time today, and I feel indifferent about this), and (4) does not need risks to be present (e.g., no safety consequences for me if the ferry is not on time).

We argue that although confidence is only one characteristic of trust, it is the most important ingredient for trusting correctly. Evidence, knowledge, skills, and experience used as grounds for justified confidence can be extrapolated into fostering correct trust. In the previous example, Bob is basing his trust on Carol's safety explanations of the ferry. He extrapolates the evidence to acquire new knowledge and builds up his confidence. Using his own reasoning and motivation, he integrates all the information and concludes that he has enough trust to give the ferry a try. This means that he will still judge whether the ferry is really to be trusted by observing, for example, the ferry's safety mechanisms and his own experience being onboard.

## 5. Different trust needs and risks

In our use case, although the three stakeholder groups (first-hand, second-hand, and third-hand stakeholders) share the same interest (for the ferry operate successfully), their trustees and risks associated with trusting differ greatly. The stakeholder groups do not share the same risks due to the level of impact a stakeholder will experience if an incident or a near miss happens to the ferry. Consequently, the groups have different trustees. The EoT approach focuses on meeting each stakeholder group's trust needs and managing associated risks. See **Table 1** for the simplified overview of the differences in trust needs and perceived risks of the three stakeholder groups.

**Table 1** The simplified overview of different trust needs and risks of the three stakeholder groups

|  | **Main trustors** | **Main trustees** | **Examples of trust needs** | **Examples of risks** |
|---|---|---|---|---|
| **First-hand stakeholder group** | • Passengers (main stakeholders)<br>• Ferry operators<br>• Waterborne actors<br>• Other waterborne vehicles | • Second- and third-hand stakeholder groups | • Safety<br>• Security<br>• Comfort<br>• A better transport option | Injury, death, emotional distress |
| **Second-hand stakeholder group** | • Manufacturers<br>• Vendors or suppliers | • Technology<br>• Each other<br>• Third-party vendors or suppliers<br>• First-hand stakeholder group | • First-hand stakeholders use the ferry as intended<br>• Technology functions as anticipated<br>• Third-party vendors or suppliers provide conditions of products as claimed<br>• Development teams provide outcomes as claimed | • Liability<br>• Public responsibility |
| **Third-hand stakeholder group** | • Regulators<br>• Investors<br>• Insurance<br>• Other funders | • Second-hand stakeholder group<br>• Each other<br>• First-hand stakeholder group (to some extent) | • Regulators create relevant and effective regulations<br>• Investors create an accurate cost-benefit analysis<br>• Manufacturers comply to relevant regulations<br>• Operators operate the ferry as intended and approved | • Reputation loss<br>• Monetary loss |

First-hand stakeholders, especially passengers, have the greatest risk. Thus, their trust in the safety of the ferry is of the utmost importance in the ecosystem. The risk of an incident, a near miss, or even an accident does not affect the personal safety of onshore or remote operators, but they are on the frontline of the safety of the ferry. Although the type of risk is different from the passengers', the impact of an incident, a near miss, or an accident will affect onshore and remote operators as significantly.

In contrast, the second-hand stakeholder group holds the biggest responsibility in making sure that the ferry is fit for public operation. The risk is lower for the third-hand stakeholder group if the regulators create relevant and effective regulations and if the investors make an accurate cost-benefit analysis. Trust between the second- and third-hand stakeholders needs to be achieved as gaps between them can lead to otherwise preventable incidents. For example, the regulations should be fit-for-purpose for the ferry and the ferry must be manufactured and operated according to the regulations and other relevant safety standards, guidelines, and approaches. A mismatch in this can lead to, for example, loopholes and blind spots in the safety of the ferry which directly affects the passengers in specific and public safety in a broader sense.

## 5.1. Passengers' other trust needs and perceived risks

The ferry is intended to be a transport for all and wishes to reflect inclusivity in its operation. In an internal report, Pantelatos and St.Clair (2022) conducted a study on passenger engagement and found that the study participants, i.e. potential passengers of the use case ferry, were concerned with not only safety but also security. They worried about crimes, violance, and vandalism onboard as a result of lack of authority figure on the unmanned, fully autonomous ferry. These concerns led to the discussion of the importance of camera surveillance systems onboard as well as the possibility for those onboard to contact an onshore or remote operator to report any concern.

This suggested solution may need additional explanations of the security of the surveillance systems with regards to cyberattacks and privacy. After all, the ferry is a public space and any surveillance systems must comply with relevant regulations related to privacy. Future work can be directed to answer the question on how public transport, like a ferry and other autonomous systems to be used by public, balance the need of security and privacy.

Another relevant topic discussed by participants in Pantelatos and St.Clair's study was how the ferry can be inclusive for all passengers including those in more vulnerable groups such as children, blind or deaf people, wheelchair users, or others who need accommodations. Some solutions were suggested such as safety markings on the ground, use of both audio and visual communications, braille, wheel-chair accessible pathways, prioritised seating areas, and child-proof onboard design. More research is needed to enable inclusivity for the passengers.

## 6. Communicating grounds for justified confidence

Trusting the ferry itself is not enough to make the ferry system truly trustworthy (Jacovi et al., 2021). As in the example, Bob is still curious whether the ferry really delivers as he expects. Importantly, any information about the ferry, such as its safety mechanisms, should be communicated in the language that target users understand and will interpret as intended in order to help them make informed decisions.

Glomsrud et al. (2019) state that explanations are a human need, and that explainable AI is one effective method to make the complexity of autonomous systems understandable to their target users. Felzmann et al. (2020) suggest a guidance to concretise Transparency by Design by, among other methods, optimising communication with different stakeholders and embedding relatable and concrete measures into design and implementation. Pantelatos et al. (2023) and Pantelatos and St. Clair (2022) have investigated how the passengers felt on our autonomous ferry use case in a trial study. The study findings show the importance of ensuring that the passengers of the ferry understand the ferry's behaviours.

Confusion about the ferry's behaviours, for example, when it suddenly stopped on the water without a clear reason, gave the passengers uncertainty and highlighted the passengers' strong desire to receive adequate information on what is happening at all times while on the ferry (Pantelatos et al., 2023; Pantelatos & St.Clair, 2022). Communicating the mechanisms of the ferry is one important step, but it is even more important to ensure that the receivers understand the information accurately and that receiving this information has the intended positive effect, such as reducing their uncertainties and concerns (Felzmann et al., 2019). It is thus important that the information being communicated, and the expectations of the ferry are consistent with reality (Jacovi et al., 2021; Pantelatos et al., 2023).

Back to our example, when Bob boards the autonomous ferry, he feels that the ride is not only enjoyable but also makes him feel safe as a passenger. He feels that the safety mechanisms he learned about earlier from Carol are being put into practice. He can see a safety certificate on the ferry, he receives adequate information about the operational safety of the ferry through brochures and posters onboard, emergency plans and tools are in place, and when the waves become bigger and affect the ferry, a remote operator informs the passengers that this is normal and soon the ride will be calmer and that they have full control of the ferry, which is exactly what happens next.

Bob arrives at the destination safe and sound as he expects. Bob has trusted the ferry correctly, and the ferry has demonstrated that Bob's confidence is justified through the mechanisms that are being practiced. Bob's first-hand experience and newly acquired knowledge about the ferry's safety mechanisms have become the grounds for him to convince Alice that the ferry is in fact safe, inviting Alice to participate in the ecosystem of the ferry. If Alice has equal confidence in Bob, she will use these grounds to justify her confidence in Bob's judgement and challenge her initial judgement towards the ferry.

## 7. Discussion

Assurance plays three key roles in enabling the EoT approach. First, assurance can help foster correct trust. Assurance is different from merely confidence. Assurance implies that evidence or knowledge used as grounds for justified confidence is collected, analysed, interpreted, and communicated in *a systematic, targeted, and meaningful way*. An assurance process should be completed using a holistic point of view (i.e., a systems perspective (Haugen, 2022)) including identifying target stakeholders and their pains, gains, and motivation. Assurance must be compliant with regulations and follow relevant standards, guidelines, and/or approaches.

It must also adequately communicate the assurance mechanisms and results to the target stakeholders. This communication should be tailored to different stakeholder groups because they are likely to have different trust needs and risk perceptions. Bob, for example, has his own trust process that is likely different from that of Alice. Assurance should be able to provide satisfying confidence to Bob, Alice, and other stakeholders equally. Although assurance cannot guarantee trust, it builds a very strong foundation. We argue that assurance is the only proven and available way to help foster and maintain the correct trust in an EoT.

Second, assurance can help overcome the fact that trust is not only situational and context-dependent (Bach et al., 2022), but also changeable over time (Elkins & Derrick, 2013). These trust characteristics can be a challenge for an ecosystem which cannot function optimally unless the correct trust is present at all times, or at least during interactions among and between the stakeholders and technology. Assurance mechanisms should be created with involvement from all relevant stakeholder groups to ensure fit-for-purpose solutions and reduce stakeholder uncertainties and concerns at crucial moments (e.g., when another ferry is coming toward, or when there is a big wave or bad weather).

Identifying the crucial moments, specific trust needs, and perceived risks of the three stakeholder groups will allow the creation of assurance mechanisms that should reduce the negative consequences of trust characteristics (i.e., situational, context-dependent, and changeable). For Bob and Alice, one of the crucial moments onboard is when there are big waves coming to the ferry. Identifying this could lead to the creation of an assurance mechanism, for example, one that requires communicating to the passengers what is happening and the strategy to deal with the waves, including the predicted timeline of the waves.

Another crucial moment is when two ferries are in the same pathway. An assurance mechanism could instruct the autonomous ferry to keep distance, slow down, move away, or even stop, while communicating the reasoning for the ferry's behaviours to the passengers (Pantelatos & St.Clair, 2022). An assurance process thus can identify specific trust needs and perceived risks so that the appropriate information and solutions can be communicated to the target stakeholders.

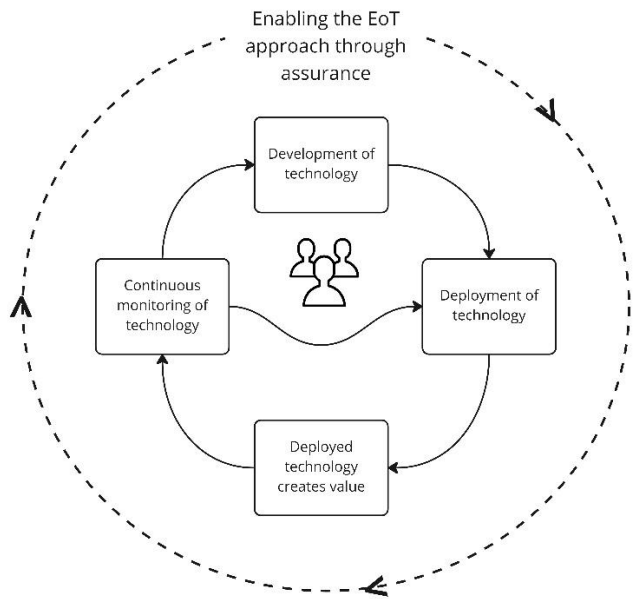

**Fig. 4** Enabling the EoT approach through assurance mechanisms from the development phase, deployment, through to continuous monitoring of technology using an iterative process

Third, enabling the EoT approach through assurance is an iterative process from development, to deployment, to value creation as a result of the deployment, through to continuous monitoring of the deployed autonomous system (**Fig. 4**). Stakeholders are at the centre of the iterative process because their feedback is fundamental to the continuation of the operation of autonomous systems. Following Heraclitus' quote "change is the only constant in life", it is almost unrealistic to expect that the only change in an ecosystem is the deployment of technology. Even if the deployment of technology makes the biggest ripple effect to the ecosystem, the original deployment strategy may not be relevant over time and will need continual adjustment through continuous monitoring.

As another example, automation complacency and bias that may occur over time as autonomous systems become embedded into society (Dixon, 2020; Rodriguez et al., 2019) should be monitored to prevent undesirable consequences (Aroyo et al., 2021). This changing dynamic of ecosystems is one reason why monitoring should be continuous, using more frequent checks and/or real-time data (e.g. Dodero et al., 2021; Minkkinen et al., 2022; Yin et al., 2019). Accordingly, deployment cannot just be a one-off activity; outcomes of deployment should be used as feedback to continuously improve and adjust the technology and deployment strategy. This may mean starting slowly. For example, there may initially be an operator onboard who is gradually phased out as more data, experiences, and feedback are gathered.

## 8. Future work and conclusion

We focused on an autonomous system in this paper, but the EoT approach can be applied in almost every context where complex and interdependent stakeholders, systems, and technologies want to create value in the presence of risk in a natural environment. The scope of the approach can follow according to each use case's purpose. The approach can, for example, be applied for sustainable business models to identify the interdependent actors, their trust needs, and expected value. Another example of how the EoT approach can be applied is in a hospital department to investigate trust needs of different stakeholder groups (patients, clinicians, and vendors) for achieving patient safety. Future work should explore more use cases to investigate how the EoT approach can foster trust in deployment of technologies in different contexts and industries.

The EoT approach through assurance is one tool to ensure value creation in the presence of risk. There are other tools, methods, and approaches with similar interest. For example, technology acceptance models (Marangunić & Granić, 2015), human-computer interactions (e.g. Carmona et al., 2021), sociotechnical systems (e.g. Sony & Naik, 2020), and other holistic approaches (e.g. Lewandowski, 2016; van Gemert-Pijnen et al., 2011). There is a need to ensure that technology deployment is successful and future work should focus on understanding which methods and approaches fit best in which situations. In addition, future work needs to further investigate the interactions between available tools, methods, and approaches and which to combine to reduce their limitations and enable their strengths.

Although the EoT approach employs a holistic approach, there will still be unknown-knowns and unknown-unknowns. For example, interdependencies between all stakeholders, elements, and systems in an ecosystem may have a much bigger and wider impact than originally anticipated. Another example is the real impact of Generative AI over time.

Another consideration is the scope of the EoT approach. Since these interdependencies can be as wide as the horizon, where should we put a border in an ecosystem? Can the ripple effect be used to track interdependencies and determine the scope? Even a use case like the one presented in this paper may include a much bigger and wider ecosystem than originally thought. How could we onboard new stakeholders who enter the ecosystem with a mindset opposed to the stakeholders in the ecosystem (e.g., new competitors to existing businesses)?

Since the main focus of the EoT approach is about trust, more work is needed to understand how trust interacts with other equally important factors such as regulations, culture, and diversity. The EoT approach may lead to the desired result if things go well, showing the gap in concretising and operationalising the EoT approach and how to foster and maintain trust over time as well as recovering broken trust in an ecosystem.

In conclusion, successfully deploying autonomous systems requires multiple factors to align. The EoT approach can be a departure point for creating a suitable deployment strategy. Because the nature of such deployment can be rather disruptive to an existing ecosystem, assurance is a prerequisite to ensure that risk is managed, trust is shared, and value is created.